\documentclass[twocolumn,aps,floatfix,showpacs,pra]{revtex4}
\usepackage{color}
\usepackage{amsmath}
\usepackage{graphicx}
\usepackage{psfrag}
\usepackage[normalem]{ulem}

\begin{document}
\title{Criticality in the Bose-Hubbard model with three-body repulsion}
\author{Tomasz Sowi\'nski$^{1}$, Ravindra W. Chhajlany$^{2,3}$, Omjyoti Dutta$^{4}$, Luca Tagliacozzo$^{5}$, and Maciej Lewenstein$^{2,5}$}
\affiliation{
\mbox{$^1$Institute of Physics of the Polish Academy of Sciences, Al. Lotnik\'ow 32/46, 02-668 Warszawa, Poland}\\
\mbox{$^2$ ICFO - The Institute of Photonic Sciences, Av. Carl Friedrich Gauss, num. 3, 08860 Castelldefels (Barcelona), Spain} \\
\mbox{$^3$ Faculty of Physics, Adam Mickiewicz University, Umultowska 85, 61-614, Pozna{\' n}, Poland} \\
\mbox{$^4$ Instytut Fizyki imienia Mariana Smoluchowskiego, Uniwersytet Jagiello\'nski, \L ojasiewicza 11, 30-348 Krak\'ow, Poland} \\
{$^5$ ICREA - Instituci\'o Catalana de Recerca i Estudis Avan\c cats, 08010 Barcelona, Spain}
}
\date{\today}

\begin{abstract}
We study the attractive Bose-Hubbard model with a tunable, on-site three-body constraint. 
It is shown that the critical behavior of the system undergoing a phase transition from pair-superfluid to superfluid at unit filling depends on the value of the three-body repulsion. In particular, we calculate critical exponents and the central charge governing the quantum phase transition. 
\end{abstract}
\maketitle

\section{Introduction}
Amazing experimental progress on ultra-cold atoms confined in optical lattices has generated an almost ideal arena for simulating and testing various Hubbard-like models \cite{Zoller1,Zwerger,Lewenstein,DuttaRev}. Since the first experimental confirmation of the quantum phase transition (QPT) from the Mott insulator phase to the superfluid (SF) phase in the Bose-Hubbard model  \cite{Greiner},   many extensions of the model have been proposed theoretically \cite{Zwerger,Lewenstein}. Extended  Hubbard models (for fermions and bosons) take into account processes induced by the internal structure of interacting particles, long-range forces, and higher bands of optical lattices. All of these lead directly to additional interaction terms in the Hamiltonian which support a rich plethora of physical properties \cite{DuttaRev,LewensteinBook} awaiting experimental verification.  

The availability of such systems in less than three dimensions promises avenues to investigate the nature of phase transitions in low dimensionality. Particularly, for one-dimensional 1D quantum systems it opens a new route to look for and  test conformally invariant systems with different central charges ${\cal C}$. It is known that for systems with ${\cal C}<1$ the critical exponents are fully determined by the charge, and they they do not depend on the details of the microscopic model \cite{Cardy}. This is a specific manifestation of the so-called {\em universality hypothesis} close to quantum phase transitions. 
However there are models  violating the hypothesis in the sense that their critical exponents  vary continuously with respect to the microscopic parameters of the model. Prominent examples of such non-universal behavior are found in the Ashkin-Teller model  \cite{Ashkin,Potts,Kadanoff}, which is a specific instance of a compactified free bosonic model generically described by Tomonaga-Luttinger liquid theory \cite{RigolRMP}, the eight-vertex model solved exactly by Baxter \cite{Baxter}, arbitrary spin Heisenberg models \cite{Alcaraz},  etc. From the conformal invariant point of view these are related to  field theoretical models with central charge ${\cal C}\geq 1$. 

A recent unconventional idea leading to a system described by a CFT with a fractional central charge ${\cal C}>1$ than one  centered around the study of attractive ultra-cold bosons described by a Bose-Hubbard model with an on-site three-body constraint \cite{Daley,Mark}. The three-body constraint dis-allows  the occupation of a site by three or more bosons and is equivalent to an infinite on-site three-body repulsion term. Importantly, the constraint  inhibits the collapse of bosons in the system on to a single site and the resulting competition between tunneling and attractive two-body interactions leads to interesting quantum critical behavior. At low tunneling, a pair-superfluid (PSF) phase is supported, while at high tunnelings the more conventional single-particle SF phase is favored. These two phases are separated, at unit filling, by an exotic Ising type quantum phase transition  characterized by a  $U(1)\times Z_2$ conformal field theory with fractional central charge ${\cal C}=1+1/2$. 
 This model is  experimentally feasible, as the three-body constraint can be enforced by rapid three-body recombination processes corresponding to the decay into the continuum of unbound states \cite{Daley} and thus loss of particles from the underlying optical lattices. This has driven interest in deep studies of the model \cite{Lee,Diehl1,Diehl2,Chen,Bonnes,Ng}. In contrast to other systems explored with the PSF phase \cite{Schmidt,Arguelles,Zhou,Trefzger,Sowinski1}, here the PSF emerges through local interactions only. It was further shown that such a model can be mimicked by using a unit filling Mott insulator of spin-1 atoms \cite{Mazza}.

\section{The Model}
In this article, we consider the attractive Bose-Hubbard model with relaxing of the three-body on-site repulsion from the hard-core limit to a finite value. We assume that the  finite three-body repulsion is still sufficiently large to prevent the system from collapsing. While the three-body losses dominate in the system, they are not instantaneous. We restrict ourselves to the 1D lattice wherein the system of $L$  sites (with open boundary conditions) is described by the extended Bose-Hubbard Hamiltonian:
\begin{align}
H &= -J \sum_{i=1}^{L-1}  (\hat{a}_i^\dagger \hat{a}_{i+1}+ \hat{a}_{i+1}^\dagger \hat{a}_{i}) + \frac{U}{2}\sum_{i=1}^{L} \hat{n}_i(\hat{n}_i-1) \nonumber \\
&+ \frac{W}{6}\sum_{i=1}^{L} \hat{n}_i(\hat{n}_i-1)(\hat{n}_i-2). \label{Hamiltonian}
\end{align} 
The $\hat{a}_i$ is the annihilation operator of a single boson at site $i$ and $\hat{n}_i = \hat{a}_i^\dagger\hat{a}_i$ is the local particle number operator. The two-body interaction $U<0$ and $W$ is the repulsive three-body interaction parameter. This three-body term can be understood as an effective way of taking into account higher orbital physics and three-body losses. In the limit $W\rightarrow \infty$ the last term in the Hamiltonian is equivalent to the three-body on site hard-core constraint $\forall_i (a^\dagger_i)^3 \equiv 0$ introduced and studied in detail earlier \cite{Lee,Diehl1,Diehl2,Chen,Bonnes,Ng}. Here, we examine the properties of the QPTs the system undergoes when the three-body term is tunable and large but finite. While populating a lattice site with three bosons  is not completely forbidden, the associated energy cost makes this process unfavorable. Thus states with sites occupied  by more than two bosons affect the system properties mainly via virtual processes. It is worth noticing, that such a model was also studied in the opposite case, i.e., for repulsive two-body interactions and attractive three-body ones, where contraction of insulting lobes was predicted \cite{Silva,Sowinski2,Safavi,Ejima,Silva2,SowinskiRavi,Sowinski3}.

\section{Exact diagonalization approach}

We study the QPT numerically using two approaches: Exact Diagonalization of small clusters of size up to $L=12$ sites and the Density Matrix Renormalization Group method (DMRG) for a much larger size of $L=128$. Using these methods, we study the variation with the repulsion parameter $W$ of quantum critical properties at the PSF-SF quantum phase transitions.

\begin{figure}
\centering
\includegraphics{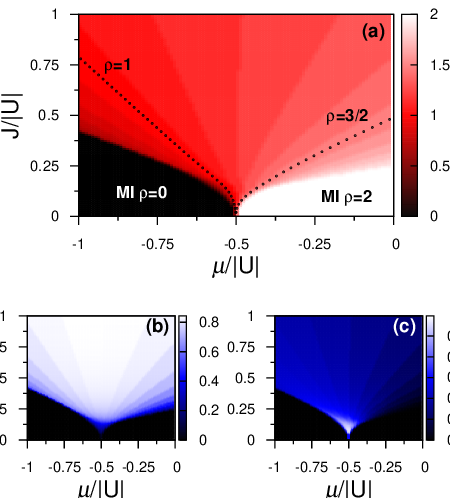}
\caption{(color on-line) Phase diagram of the hard-core model ($W\rightarrow \infty$) obtained using exact diagonalization in the full many-body basis on a $L=8$ site lattice. (a) Average filling $\rho$; (b) single-particle superfluid hopping $\phi$; (c) pair-superfluid (PSF) hopping $\Phi$. For $\rho=0$ and $\rho=2$ the system is always in the insulating phase. For intermediate filling, depending on the tunneling, the system is either in the SF or in the PSF phase. }  \label{Fig1}
\end{figure}

It is known that properties of Hubbard systems crucially depend on the relation between the total number of particles $N$ in the system  and the number of lattice sites $L$. Therefore, it is convenient to introduce the filling factor $\rho=N/L$.  One can also
reformulate the many body problem in the grand canonical ensemble by 
adding the term $-\mu \sum_i\hat{n}_i$ to the Hamiltonian $H$. In this picture, the average number of particles in the system is determined by the chemical potential $\mu$. In Fig. \ref{Fig1}, the phase diagram of the hard-core system ($W \rightarrow \infty$) is presented. It is worth emphasizing that although the Hamiltonian is mathematically well defined for all fillings $\rho\leq 2$, it describes the physical system correctly only for low fillings. At higher fillings the influence of higher orbitals should be taken into account for accuracy. In the following, we focus on the unit filling case $\rho=1$, i.e. when the total number of particles $N$ is equal to the total number of lattice sites $L$. 
 
The hard-core model with $W\rightarrow \infty$ in \eqref{Hamiltonian} undergoes the phase transition from PSF to SF for any $\rho<2$. At the mean-field level, which yields a qualitative picture above $d=1$, the SF phase is characterized by non vanishing order parameters $\langle a_i\rangle\neq 0$ and $\langle a_i^2\rangle\neq 0$ in the grand canonical ensemble. In contrast, in the PSF phase the condensate fraction $\langle a_i\rangle$ vanishes. In particular, the two phases differ under the transformation w.r.t. the $Z_2$ (Ising) symmetry $a \rightarrow - a$,\textit{ viz.}, the PSF is invariant (corresponding to the disordered Ising phase), while the SF breaks this symmetry (corresponding to the ordered Ising phase).
 Note that the difference between the PSF phase and insulating phases (which we do not consider further) lies in the dimer order parameter $\langle a_i^2 \rangle$, which in the PSF remains non vanishing, {\it i.e.}  non-local correlations still exist.
In $d=1$, on the other hand, quantum fluctuations inhibit the formation of true long-range order with explicit breaking of the continuous $U(1)$ symmetry. In this case, however, quasi-long-range order characterizes the superfluid phases with algebraically (slowly) decaying long range correlations (Indeed, quantum-classical equivalence allows us to think of the system as a (1+1)-dimensional thermal classical system). The  SF-PSF transition we discuss below concerns the transition between these two quasi-long range orders.
While the two particle correlation function $\langle (a^{\dagger}_i)^2 (a_{i+j})^2\rangle$ decays algebraically in both phases, the single particle function $\langle a^{\dagger}_i a_{i+j}\rangle$ decays polynomially in the SF phase and exponentially in the PSF phase.

The nearest-neighbor correlation functions, or so-called hopping fields \cite{Sowinski1} $\phi = \sum \langle a_{i+1}^\dagger a_i \rangle/(L-1)$ and $\Phi = \sum \langle a_{i+1}^\dagger a_{i+1}^\dagger a_i a_i\rangle/(L-1)$  already indicate the changes in superfluid tendencies in finite  systems.
\begin{figure}
\centering
\includegraphics{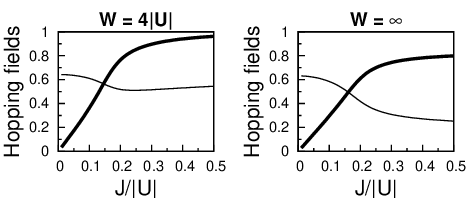}
\caption{Hopping fields $\phi$ and $\Phi$ obtained with the help of exact diagonalization of Hamiltonian \eqref{Hamiltonian} at unit filling in a lattice of size $L=10$ for two extreme choices of three-body repulsion $W$. The PSF hopping (thin line) is always finite and it dominates for low tunneling rates $J$. The SF hopping is negligible for low tunneling rates but dominates for high values. For some intermediate tunneling value,  the system undergoes a crossover between these two superfluid phases, which in the thermodynamic limit will turn into a true quantum phase transition. }  \label{Fig2}
\end{figure}
To examine the critical behavior of the system for $\rho=1$ for different values of three-body repulsion, we first perform exact diagonalization of the Hamiltonian for a fixed lattice size $L$. In this way, for a given tunneling $J$ (measured in the units $U=1$) we find the exact ground state of the system $|\mathtt{G}(J)\rangle$ and its energy $E_G(J)$. It is a straightforward observation that depending on the tunneling $J$, the system exhibits non-local correlations. For low tunneling rates, SF hopping $\phi$ is negligible, and PSF hopping $\Phi$ dominates in the system (see Fig. \ref{Fig2}). For higher tunneling amplitudes the SF hopping increases and plays a crucial role in determining the properties of the system. Apart from some details this general picture holds also even for very soft three-body repulsions [see Fig. \ref{Fig2}b]. Hence the QPT from PSF to SF exists for any reasonable value of $W$. 

We now use the fidelity susceptibility (FS) as a quantitative indicator of the PSF-SF QPT, defined as  \cite{Gu,Cozzini,You,Buonsante,Rigol1,Rigol2}:
\begin{align}
\chi(J)=-\left.\frac{\partial^2{\cal F}(J,\delta)}{\partial \delta^2}\right|_{\delta=0} = -2\lim_{\delta\rightarrow 0}\frac{\ln {\cal F}(J,\delta)}{\delta^2}
\end{align}
where ${\cal F}(J,\delta)=|\langle G(J)|G(J+\delta)\rangle|$ is the fidelity of neighboring ground states of the system differing by a small change in the tunneling. In the thermodynamic limit, i.e. when $L \rightarrow \infty$, the singular behavior of FS is an efficient indicator of the phase transition of finite rank. For finite system size, the FS is a smooth function of $J$ but some non monotonic behavior is still present. In the left panel in Fig. \ref{Fig4}, the FS for two three-body repulsions $W$  are presented for different system sizes $L$. One finds that the maximal amplitude of the FS, as well as its position crucially depend on the system size $L$. The observed behavior of the FS suggests that it will diverge for an infinite size system (while the fidelity $\cal F$ will collapse to 0). Therefore, the studied QPT falls under the paradigm of Landau type phase transitions and can be well understood within the phenomenological Landau symmetry-breaking theory \cite{You,Zenardi,Liu,Cozzini2,Venuti,Buonsante}. This picture is consistent with previous results for the hard-core model, since it is known that the PSF-SF transition it undergoes is of the Ising-type at unit filling $\rho=1$ \cite{Diehl1,Diehl2,Lee,Bonnes}.

From this behavior of the FS, by using the {\em finite scaling method} (see, for example, \cite{Gu,Newman}), one can extract the position of the critical tunneling $J_c$ and critical exponents describing divergences of system properties near the critical point. To do this systematically, we introduce the reduced tunneling $t = (J-J_c)/J_c$ as the measure of distance from the critical point. Then we assume that in the neighborhood of the critical point the correlation length $\xi$ and the FS $\chi$ diverges according to the power laws $\xi \sim |t|^{-\nu}$ and $\chi \sim |t|^{-\gamma}$. Combining these two relations, we find the connection between the FS and the correlation length for an infinite system size, {\it viz.} $\chi \sim \xi^{\gamma/\nu}$. For finite-size systems the situation is different since the correlation length $\xi$ cannot be larger than the system size $L$. The above scaling relation is valid only as long as $\xi \ll L$. When $\xi$ become comparable to the system size $L$, then the FS approaches some finite value and does not change. This observation can be expressed by introducing an additional function $\chi_0(x)$ which has some asymptotic value $K$ in the limit $x\rightarrow \infty$ and diverges as $x^{\gamma/\nu}$ when $x \rightarrow 0$. Then  
$\chi = \xi^{\gamma/\nu} \chi_0(L/\xi)$. To make this relation independent of the correlation length, we  finally introduce {\em the scaling function} $\tilde{\chi}(x) = x^{-\gamma} \chi_0(x^\nu)$. In this way, we can write the susceptibility as a function of the reduced tunneling $t$ and system size $L$
\begin{align}
\chi(t,L) = L^{\gamma/\nu} \tilde{\chi}(L^{1/\nu}t). \label{EqChi}
\end{align}
The scaling function $\tilde{\chi}$ is analytical and, in principle,  is a universal function independent of the system size. In practice, it contains some dependence on system size away from its extrema.

\begin{figure}
\centering
\includegraphics{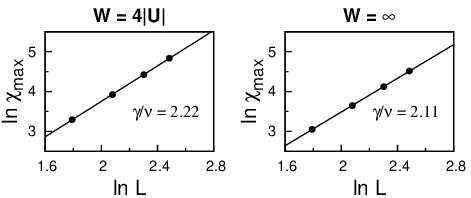}
\caption{Maximal value of the fidelity susceptibility $\chi_{\mathrm{max}}$ as a function of the system size $L$ for two choices of three-body repulsion values. On a log-log scale the data points fit the linear predictions of the finite-scaling hypothesis. The slope corresponds to the ratio of the two critical exponents $\gamma/\nu$.}  \label{Fig3}
\end{figure}

\begin{figure}
\centering
\includegraphics{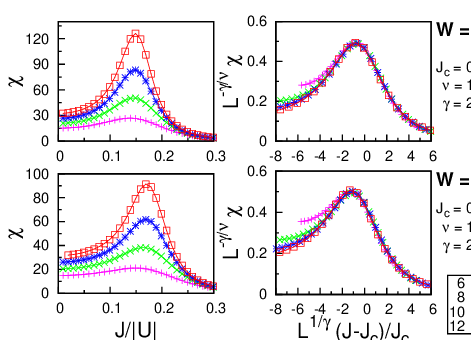}
\caption{(color on-line) Fidelity susceptibility (FS) $\chi$ as a function of tunneling amplitude $J$ for two values of the three-body repulsion ($W=4|U|$ and $W=\infty$) obtained for different system sizes $L=6, 8, 10, 12$. (left panel) FS is a non-monotonic function of $J$ and its peaks are indicators of changes in the ground state of the system (quantum phase transitions). (right panel) the same plot with rescaled variables. Rescaling parameters (critical tunneling $J_c$ and critical exponents $\nu$ and $\gamma$) for which all curves collapse to one universal curve are given at the right. The critical exponent $\nu$ is insensitive to the strength of the three-body repulsion. In contrast, the critical exponent $\gamma$ as well as the critical tunneling $J_c$ depend on $W$. } \label{Fig4}
\end{figure}

Equation \eqref{EqChi} can be used to determine the critical tunneling $J_c$ and critical exponents $\nu$ and $\gamma$. For example, there is a very simple way of determining the relation between the two critical exponents $\gamma$ and $\nu$. By plotting the maximal amplitude of the fidelity susceptibility $\chi_{\mathrm{max}}$ as a function of the system size $L$ on a log-log scale, one expects a linear relation with a slope of $\gamma/\nu$. The plots in Fig. \ref{Fig3} show that even for small systems, the linear regression is almost perfectly satisfied for the hard-core model ($W\rightarrow \infty$) as well as for very soft three-body repulsion ($W=4|U|$). Absolute values of the critical exponents as well as the value of the critical tunneling $J_c$ can be found via the {\em data collapse} method, as explained below. If all critical parameters were known and indeed the relation \eqref{EqChi} were fulfilled, then plotting $L^{-\gamma/\nu} \chi$ versus $L^{1/\gamma} t$ for different system sizes would cause all curves to collapse to the universal function $\tilde{\chi}$. Importantly, this data collapse occurs only for a unique set of critical parameters. Therefore, in principle all critical parameters can be determined by finding proper rescaling so that all data for different system sizes fall on to one curve.  Examples of  {\em data collapse} for the studied system in the two cases $W=4|U|$ and $W=\infty$ are presented in the right panel in Fig. \ref{Fig4}.
In Table \ref{Tabela}. the critical parameters for intermediate three-body interactions $W$ are listed. 
These results show that, as expected, the critical point corresponding to the transition from PSF to the normal SF phase depends on the three-body repulsion. Interestingly, however, in addition, we find that the critical exponent $\gamma$ varies with $W$ and is thus dependent on the microscopic details of the model. This strongly suggests that Hamiltonian (\ref{Hamiltonian}) violates the universality hypothesis.

\begin{table} 
\begin{tabular}{c|ccc}
\hline
\hspace{0.2cm}$W/|U|$\hspace{0.2cm} & \hspace{0.2cm}$J_c/|U|$\hspace{0.2cm} & \hspace{0.3cm}$\gamma$\hspace{0.3cm} & \hspace{0.3cm}$\nu$\hspace{0.3cm} \\
\hline \hline
4 & 0.160 & 2.23 & 1.0 \\
5 & 0.170 & 2.18 & 1.0 \\
6 & 0.175 & 2.16 & 1.0 \\
8 & 0.180 & 2.14 & 1.0 \\
10 & 0.180 & 2.13 & 1.0 \\
15 & 0.185 & 2.12 & 1.0 \\
20 & 0.185 & 2.11 & 1.0 \\
100 & 0.190 & 2.11 & 1.0 \\
$\infty$ & 0.190 & 2.11 & 1.0 \\
\hline
\end{tabular}
\caption{Critical parameters of the studied system for different values of the three-body interaction $W$. Increasing the three-body repulsions leads to enhanced stability of the PSF phase ({\it i.e.}  the critical tunneling $J_c$ increases) while the critical exponent describing divergences of the fidelity susceptibility ($\gamma$) decreases. Results are obtained by data collapse of finite-size results. \label{Tabela}}
\end{table}

\section{Entanglement entropy approach}

In the framework of renormalization group theory, the numerical results above are compatible with the presence of a line of conformally invariant fixed points governed by a marginal operator with some overlap with the three body repulsion term.
We can further check this scenario by considering the scaling of the entanglement entropy (EE) of a block of contiguous sites with its growing size. If the system is conformally invariant, the prefactor to the scaling provides an estimate of the central charge $\cal C$ of the CFT underlying the PSF-SF quantum phase transition for a given repulsion $W$.  For systems with open boundary conditions, the EE of a block of size $l$ on a finite lattice of length $L$ takes the form \cite{Lafro,Vidal,Latorre,Holzhey,Calabrese,Calabrese2,Bhaseen}: 
\begin{align} 
{\cal S}_\pm(x) = \frac{\cal C}{6}\ln\left[\sin\left(\frac{\pi l}{L}\right)\right] + s_\pm(L) + {\cal O}\left(\frac{l}{L}\right),
\label{eescaling}
\end{align}  
including an alternating term $s_\pm$ (depending on the parity of the block sizes), arising due to the open boundary conditions as well as other corrections of order ${\cal O}(l/L)$ including a boundary induced term. 

\begin{figure}
\centering
\includegraphics{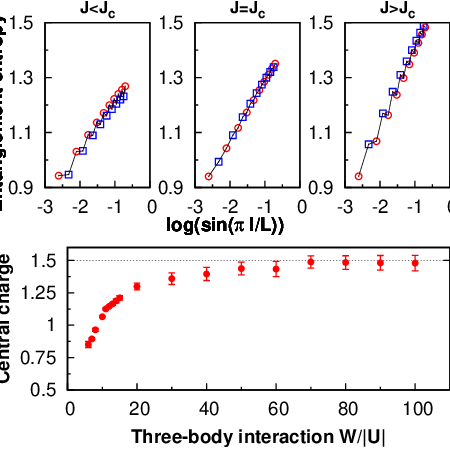}
\caption{(color on-line) (upper panel) Scaling of the entanglement entropy \eqref{eescaling} as a function of the scaled block size for an exemplary case $W/|U|=30$. The QPT point corresponds to the vanishing of the oscillatory dependence on the parity of the block size. (bottom panel) The central charge of the system at the SF-PSF transition for different values of three-body interactions. } \label{Fig5}
\end{figure}

We calculate the EE scaling under open boundary conditions using the DMRG method for system size of $L=128$ and consider its behavior versus the first term in \eqref{eescaling}. Deep in the single particle SF phase, there is a simple oscillating dependence on the parity of the block size as seen for an exemplary value of $W/|U|=30$, although the scaling dependence on $\ln\left[\sin(\pi l/L)\right]$ is approximately linear (Fig. \ref{Fig5}). On the other hand, the behavior for $J<J_c$ is markedly different. The entropy does not \textit{apparently} scale linearly with $\ln\left[\sin(\pi l/L)\right]$ for \textit{small block sizes}. This is possibly the result of the interplay of the combination of open boundary conditions and finite system size which seems to lead to substantial corrections in the studied attractive Bose-Hubbard model. Indeed one would expect the PSF phase with algebraically decaying correlations also to be characterized by logarithmic scaling just as the SF phase. Significantly though, the oscillatory dependence on the parity of the block length is reversed compared to the case of the single particle SF. We therefore identify the change from PSF to SF  (and thus the critical value $J_c$) with switching of the oscillating behavior in the finite-size system, which also corresponds to the best possible linearization of the entropy dependence on $\ln\left[\sin(\pi l/L)\right]$. This is true for all considered values of $W$. In particular, the critical points obtained in this manner are in agreement with those obtained via the finite-size scaling method discussed in the previous section.
Importantly, the  mentioned linearization also facilitates the direct estimation of the central charge for the QPT. The central charge thus obtained is plotted in Fig. \ref{Fig5} and shows a dependence on the repulsion $W$.

For very large $W=100$, we obtain a value very close to ${\cal C}=3/2$ as expected from 
previous results for the hard core model described by a $U(1)\times Z_2$ CFT. When $W$ is decreased, the central charge falls gradually from this asymptotic value remaining close to it for large $W$. Upon approaching the instability towards density collapse due to predominant  attractive interactions, the estimated central charge falls rapidly below the value of ${\cal C}=1$. Again, this is probably due to the interplay between open boundary conditions and finite-size effects, which make the accurate extraction of the central charge cumbersome for attractive bosons. A different, perhaps less likely, possibility is that the nature of the QPT changes as the three-body constraint is relaxed to the region prior to the system's becoming unstable towards density collapse. At the level of results obtained in this article, we are not able to exclude this possibility {\it per se} and leave this as an open problem for further research. In any case, however, as mentioned in the introduction, for ${\cal C} \geq 1$ \cite{Alcaraz} the critical exponents are not fixed by the value of the central charge and may assume continuously varying values depending on the microscopic parameter $W$ which is in agreement with results obtained in the previous section.

It is important to stress that having checked that the scaling of the EE has the same form as that predicted by CFT is not sufficient to claim that the model is conformally invariant. There are several examples of systems, albeit pertaining to disordered or long-range interactions,   known to exhibit similar scaling of the EE as displayed by CFT, even if they are not at all conformally invariant \cite{Refael,Koffel}.
In order to address the emergence of conformal invariance one should also check the scaling of the gap of the system and the form of the dispersion relation for low momenta.  We leave the issue of the presence of conformal symmetry at this QCP as an open question that deserves further studies of the model.

\section{Conclusions}

To conclude, in this article, we studied critical properties at the SF-PSF quantum phase transition in the one-dimensional attractive Bose-Hubbard model with tunable three-body constraints at unit filling. On the one hand, we obtained results for critical exponents using the standard data collapse method for small system sizes of up to $L=12$. We corroborated these results with a complementary study on a large lattice of size $L=128$, using DMRG, of the central charge associated with the phase transitions. The results presented  here provide evidence that the tunable on-site three body repulsion in the attractive Bose-Hubbard model has overlap with both a relevant and a marginal operator, in the language of Renormalization Group theory, so that it both shifts the critical point and  influences the critical properties of the PSF-SF quantum phase transition.  

Our results were obtained  for finite size systems  and as such may suffer from  finite size effects. In principle it is possible to compute the fidelity susceptibility accurately for larger systems using DMRG to check if the observed behavior survives for larger systems. One could also complement this study by investigating other quantities like the exponent in the algebraic decay of correlation
functions in real space. Similar comments concern the calculation of the central charge, which probably could be better controlled if periodic boundary conditions were used -- this type of calculation is much more computationally intensive. 
All of these possible extensions go beyond the scope of the present, already dense and numerically very involved paper. We note, moreover, that  the present calculations correspond to systems that can be studied in experiments with ultra-cold atoms and ions, which typically are far from the thermodynamic limit.  Finally let us observe that three-body hard-core bosons were shown to support hidden long-range correlations associated with gapless Pfaffian-ansatz ground states especially strongly near the Mott transition \cite{Duric}. The overlap between Pfaffian ansatz and true ground states for attractive two-body interactions diminished significantly but was still found to be substantial. Apart from the gaplessness of the two competing phases in the model studied by us, these results further serve to qualitatively illustrate the interesting strong long range correlation physics also at play in the attractive two-body model.

\section{Acknowledgements}
This research was supported by (Polish) National Science Center Grants No. DEC-2012/04/A/ST2/00090 (T.S.), No. DEC-2011/03/B/ST2/01903 (R.W.C.), and No. DEC-2012/04/A/ST2/00088 (O.D.), John Templeton Foundation, the EU IP SIQS and AQUTE, the EU Grants EQuaM and QUIC, the European Research Council (ERC AdG OSYRIS and QUAGATUA), and the Spanish Ministry project FOQUS (FIS2013-46768) and the Generalitat de Catalunya project 2014 SGR 874, Grant No. FP7-PEOPLE-2010-IIF ENGAGES 273524 (L.T.). T.S. acknowledges support from the Foundation for Polish Science (KOLUMB Programme; KOL/7/2012). R.W.C. acknowledges support for a short-term visit to ICFO from a POKL grant at the AMU and a Mobility Plus Fellowship from the Polish Ministry of Science and Higher Education.


\begin{thebibliography}{99}
\bibitem{Zoller1} D. Jaksch, C. Bruder, J.I. Cirac, C. W. Gardiner, and P. Zoller, Phys. Rev. Lett. {\bf 81}, 3108 (1998).
\bibitem{Zwerger} I. Bloch, J. Dalibard, and W. Zwerger, Rev. Mod. Phys. {\bf 80}, 885 (2008).
\bibitem{Lewenstein} M. Lewenstein et al., Adv. Phys. {\bf 56}, 243 (2007).
\bibitem{DuttaRev} O. Dutta {\em et al.}, Rep. Prog. Phys. {\bf 78}, 066001 (2015)
\bibitem{Greiner} M. Greiner, O. Mandel, T. Esslinger, T. W. H\"ansch, and I. Bloch, Nature (London) {\bf 415}, 39 (2002).
\bibitem{LewensteinBook} M. Lewenstein, A. Sanpera, V. Ahufinger, {\em Ultracold Atoms in Optical Lattices: Simulating quantum many-body systems} (Oxford University Press, Oxford, 2012).
\bibitem{Cardy} J.L. Cardy, in {\em Phase Transitions and Crititcal Phenomena}, edited by C. Domb and J.L. Lebowits (Academic, New York, 1986), Vol. 11
\bibitem{Ashkin} J. Ashkin and E. Teller, Phys. Rev. {\bf 64}, 178 (1943).
\bibitem{Potts} R.B. Potts, Mathematical Proceedings {\bf 48}, 106 (1952).
\bibitem{Kadanoff} L.P. Kadanoff, Phys. Rev. Lett. {\bf 39}, 903 (1977).
\bibitem{RigolRMP} M. A. Cazalilla, R. Citro, T. Giamarchi, E. Orignac, and M. Rigol, Rev. Mod. Phys. 83, 1405 (2011).
\bibitem{Baxter} R. Baxter, Phys. Rev. Lett. 26, 832 (1971).
\bibitem{Alcaraz} F.C. Alcaraz and M.J. Martins, Phys. Rev. Lett. {\bf 61}, 1529 (1988).
\bibitem{Daley} A.J. Daley, J.M. Taylor, S. Diehl, M. Baranov, and P. Zoller, Phys. Rev. Lett. {\bf 102}, 040402 (2009).
\bibitem{Mark} M. J. Mark {\em et al.}, Phys. Rev. Lett. {\bf 108} 215302 (2012).
\bibitem{Lee} Y.-W. Lee and M.-F. Yang, Phys. Rev. A {\bf 81}, 061604(R) (2010).
\bibitem{Diehl1} S. Diehl, M. Baranov, A. J. Daley, and P. Zoller, Phys. Rev. B {\bf 82}, 064509 (2010).
\bibitem{Diehl2} S. Diehl, M. Baranov, A. J. Daley, and P. Zoller, Phys. Rev. B {\bf 82}, 064510 (2010).
\bibitem{Chen} Y.-C. Chen, K.-K. Ng, and M.-F. Yang, Phys. Rev. B {\bf 84}, 092503 (2011).
\bibitem{Bonnes} L. Bonnes and S. Wessel, Phys. Rev. Lett. {\bf 106}, 185302 (2011).
\bibitem{Ng} K.-K. Ng and M.-F. Yang, Phys. Rev. B {\bf 83}, 100511(R) (2011).
\bibitem{Schmidt} K.P. Schmidt, J. Dorier, A. L\"auchli, and F. Mila, Phys. Rev. B {\bf 74}, 174508 (2006).
\bibitem{Arguelles} A. Arguelles and L. Santos, Phys. Rev. A {\bf 75}, 053613 (2007).
\bibitem{Zhou} X.-F. Zhou, Y.-S. Zhang, and G.-C. Guo, Phys. Rev. A {\bf 80}, 013605 (2009).
\bibitem{Trefzger} C. Trefzger, C. Menotti, and M. Lewenstein, Phys. Rev. Lett. {\bf 103}, 035304 (2009).
\bibitem{Sowinski1} T. Sowi\'nski, O. Dutta, P. Hauke, L. Tagliacozzo, and M. Lewenstein, Phys. Rev. Lett. {\bf 108}, 115301 (2012).
\bibitem{Mazza} L. Mazza, M. Rizzi, M. Lewenstein, and J.I. Cirac, Phys. Rev. A {\bf 82}, 043629 (2010).
\bibitem{Silva} J. Silva-Valencia and A. M. C. Souza, Phys. Rev. A {\bf 84}, 065601 (2011).
\bibitem{Sowinski2} T. Sowi\'nski, Phys. Rev. A {\bf 85}, 065601 (2012).
\bibitem{Safavi} A. Safavi-Naini, J. von Stecher, B. Capogrosso-Sansone, and S.T. Rittenhouse, Phys. Rev. Lett. {\bf 109}, 135302 (2012).
\bibitem{Ejima} S. Ejima, F. Lange, H. Fehske, F. Gebhard, and K. zu M\"unster, Phys. Rev. A {\bf 88}, 063625 (2013).
\bibitem{Silva2} J. Silva-Valencia and A. Souza, Eur. Phys. J. B {\bf 85}, 161 (2012).
\bibitem{SowinskiRavi} T. Sowi\'nski and R.W. Chhajlany, Phys. Scr. T, {\bf 160}, 014038 (2014).
\bibitem{Sowinski3} T. Sowi\'nski T, Cent. Eur. J. Phys. {\bf 12}, 473 (2014).
\bibitem{Gu} S.-J. Gu, H.-M. Kwok, W.-Q. Ning, and H.-Q. Lin, Phys. Rev. B {\bf 77}, 245109 (2008).
\bibitem{Cozzini} M. Cozzini, R. Ionicioiu, and P. Zanardi, Phys. Rev. B {\bf 76}, 104420 (2007).
\bibitem{You} W.-L. You, Y.-W. Li, and S.-J. Gu, Phys. Rev. E {\bf 76}, 022101 (2007).
\bibitem{Buonsante} P. Buonsante and A. Vezzani, Phys. Rev. Lett. {\bf 98}, 110601 (2007).
\bibitem{Rigol1} L. Campos Venuti, M. Cozzini, P. Buonsante, F. Massel, N. Bray-Ali, and P. Zanardi, Phys. Rev. B {\bf 78}, 115410 (2008).
\bibitem{Rigol2} J. Carrasquilla, S. R. Manmana, and M. Rigol, Phys. Rev. A {\bf 87}, 043606 (2013).
\bibitem{Zenardi} P. Zanardi and N. Paunkovic, Phys. Rev. E {\bf 74}, 031123 (2006).
\bibitem{Liu} J.-H. Liu, Q.-Q. Shi, H.-L. Wang, J. Links, and H.-Q. Zhou, e-print arXiv:0909.3031 (2009).
\bibitem{Cozzini2} M. Cozzini, P. Giorda, and P. Zanardi, Phys. Rev. B {\bf 75}, 014439 (2007).
\bibitem{Venuti} L. Campos Venuti and P. Zanardi, Phys. Rev. Lett. {\bf 99}, 095701 (2007).
\bibitem{Newman} M. E. J. Newman and G. T. Barkema, {\em Monte Carlo methods in statistical physics} (Oxford University Press, Oxford, 1999).

\bibitem{Lafro} N. Laflorencie, E.S. S\o rensen, M.-S. Chang, and I. Affleck, Phys. Rev. Lett. {\bf 96}, 100603 (2006).
\bibitem{Vidal} G. Vidal, J. I. Latorre, E. Rico, and A. Kitaev, Phys. Rev. Lett. {\bf 90}, 227902 (2003).
\bibitem{Latorre} J. I. Latorre {\em et al.}, Quantum Inf. Comput. {\bf 4},048 (2004).
\bibitem{Holzhey} C. Holzhey, F. Larsen, and F. Wilczek, Nucl. Phys. B {\bf 424}, 443 (1994).
\bibitem{Calabrese} P. Calabrese and J. Cardy, J. Stat. Mech. P04010 (2005).
\bibitem{Calabrese2} P. Calabrese, M. Campostrini, F. Essler, and B. Nienhuis, Phys. Rev. Lett. {\bf 104}, 095701 (2010).
\bibitem{Bhaseen} M. J. Bhaseen, S. Ejima, F. H. L. Essler, H. Fehske, M. Hohenadler, and B. D. Simons
Phys. Rev. A {\bf 85}, 033636 (2012).

\bibitem{Refael} G. Refael and J. E. Moore, Phys. Rev. Lett. {\bf 93}, 260602 (2004).
\bibitem{Koffel} T. Koffel, M. Lewenstein, and L. Tagliacozzo, Phys. Rev. Lett. {\bf 109}, 267203 (2012).
\bibitem{Duric} T. Duri{\'c} {\em et al.}, Arxiv:1503.05026 preprint (2015).
\end{thebibliography}
\end{document}